\documentclass[a4paper,11pt]{article}
\usepackage{pos}
\usepackage{caption}
\usepackage{subcaption}

\title{Probing AGN variability with the Cherenkov Telescope Array}

\author*[a,b]{F. Cangemi}
\author[c]{T. Hovatta}
\author[c]{E. Lindfors}
\author[b]{M. Cerruti}
\author[d]{J. Becerra-Gonzalez}
\author[e]{J. Biteau}
\author[f]{C. Boisson}
\author[g]{M. Böttcher}
\author[h]{E. de Gouveia Dal Pino}
\author[i]{D. Dorner}
\author[a]{G. Grolleron}
\author[a]{J.-P. Lenain}
\author[j]{M. Manganaro}
\author[k]{W. Max-Moerbeck}
\author[l]{P. Morris}
\author[c]{K. Nilsson}
\author[h]{L. Passos Reis}
\author[m]{P. Romano}
\author[n]{O. Sergijenko}
\author[m]{F. Tavecchio}
\author[m]{S. Vercellone}
\author[o]{S. Wagner}
\author[f]{M. Zacharias for the Cherenkov Telescope Array Consortium.}

\affiliation[a]{Sorbonne Université, CNRS/IN2P3, LPNHE, 4 place Jussieu, 75005 Paris, France}
\affiliation[a]{Université Paris-Cité, APC, 10 Rue Alice Domon et Léonie Duquet, 75013 Paris, France}
\affiliation[c]{FINCA, University of Turku, Turku, Vesilinnantie 5, 20014, Finland}
\affiliation[d]{IAC, E-38200 La Laguna, Tenerife, Spain}
\affiliation[e]{Université Paris-Saclay, CNRS/IN2P3, IJCLab, 91405, Orsay, France}
\affiliation[f]{LUTH, Observatoire de Paris, PSL Research University, CNRS, Université Paris-CIté, Meudon, France}
\affiliation[g]{CSR, North-West University, Potchefstroom 2520, South Africa}
\affiliation[h]{IAG, Aniversidade
de São Paulo, 1226 Matão Street, São Paulo, Brazil}
\affiliation[i]{Universität Würzburg, ITPA, Würzburg, Germany}
\affiliation[j]{University of Rijeka, Department of Physics, Radmile Matejčić 2, Rijeka, Croatia}
\affiliation[k]{NRAO, P.O. Box 0, Socorro, NM 87801, USA}
\affiliation[l]{Deutsches Elektronen-Synchrotron DESY, Platanenallee 6, D-15738 Zeuthen, Germany}
\affiliation[m]{INAF, Osservatorio Astronomico di Brera, Via E. Bianchi 46, I-23807, Merate, Italy}
\affiliation[n]{Astronomical Observatory, Taras Shevchenko National University of Kyiv, Kyiv, Ukraine}
\affiliation[o]{DESY, Landessternwarte, Universität Heidelberg, Königstuhl, D 69117 Heidelberg, Germany}

\emailAdd{fcangemi@lpnhe.in2p3.fr}
\emailAdd{cangemi@apc.in2p3.fr}

\abstract{Relativistic jets launched by Active Galactic Nuclei are among the most powerful particle accelerators in the Universe. The emission over the entire electromagnetic spectrum of these relativistic jets can be extremely variable with scales of variability from less than few minutes up to several years. These variability patterns, which can be very complex, contain information about the acceleration processes of the particles and the area(s) of emission. Thanks to its sensitivity, five-to twenty-times better than the current generation of Imaging Atmospheric Cherenkov Telescopes depending on energy, the Cherenkov Telescope Array will be able to follow the emission from these objects with a very accurate time sampling and over a wide spectral coverage from 20 GeV to $>20$\,TeV and thus reveal the nature of the acceleration processes at work in these objects. We will show the first results of our lightcurve simulations and long-term behavior of AGN as will be observed by CTA, based on state-of-art particle acceleration models.}

\FullConference{%
  7th Heidelberg International Symposium on High-Energy Gamma-Ray Astronomy (Gamma2022)\\
  4-8 July 2022\\
  Barcelona, Spain\\}

\begin{document}
\maketitle

\section{Introduction}
Active Galactic Nuclei (AGN) are astrophysical sources powered by accretion on supermassive black holes in galaxies and present observational signatures that cover the entire electromagnetic spectrum. 
These sources can be highly variable over the entire electromagnetic spectrum. Furthermore, variabilities are observed at different temporal scales: down to minutes for micro-variability, hours for intra-day variability and from months to years for long-term variability \citep[e.g.,][]{HovattaLindfors2019}. Rapid variabilities can be related to different processes; magneto-hydrodynamic instabilities in the disk or/and in the jets, presence of shocks or magnetic reconnection in the jets or be caused by relativistic effects due to the jet orientation \citep[e.g.,][]{Wagner&Witzel1995, Marscher2014, Camenzind&Krockenberger1992}. In any case, the details of these mechanisms are still widely debated. Studies of lightcurve variability and their associated power spectra give crucial informations on the jet dynamics, help infer the spatial scales of the emission region and provide unique insights into accelerations processes and radiative mechanisms occurring in relativistic jets.

The Cherenkov Telescope Array (CTA) will be an array of more than 50 Cherenkov Telescopes located in the northern (Canary Islands in Spain) and southern (Paranal desert in Chile) hemispheres. CTA will use three different telescope sizes referred as Large, Medium, and Small-Sized Telescope (LSTs, MSTs and SSTs). LSTs are sensitive at low energies between 20\,GeV to 200\,GeV, MSTs between 100\,GeV to 10\,TeV, whereas SSTs are sensitive to higher energies between 1\,TeV to 300\,TeV. CTA will be the largest and most advanced ground-based observatory for the detection of electromagnetic radiation in the Very High Energy (VHE) range which will make it the ideal instrument to study variability in the lightcurves of AGN at VHE.

\section{Lightcurve simulations}
	\subsection{Short term lighcurve simulations with \textsc{CtaAgnVar}}
	In order to infer CTA capabilities to characterize variability in AGN, we have developed our simulation tool called \textsc{CtaAgnVar}. \textsc{CtaAgnVar} is a python package based on \textit{Gammapy}\footnote{\url{https://gammapy.org/}} that simulates realistic observations of AGN flares. We use numerical time-dependent models as input and \textsc{CtaAgnVar} provides us with simulated lightcurves as obtained with CTA. It takes into account CTA observational constraints and source visibility during the year, uses the latest instrumental responses available for both northern and southern sites and tracks the source during the night to take into account the evolution of the elevation angle. We will use the June 2015 flare of 3C 279 to describe the different steps of our short flare simulations with \textsc{CtaAgnVar}.
	
		\subsubsection{3C 279 -- June 2015 flare}
		3C 279 is visible from both the northern and the southern sites. Therefore, lightcurve simulations have been done for both sites; however, we only show results from the north site in these proceedings, since no LSTs have been considered yet in the configuration of the southern site (and LSTs have the best sensitivity for this type of source).
		
\begin{figure}[h]
			\centering
			\begin{subfigure}[t]{0.495\textwidth}
				\includegraphics[width=\textwidth]{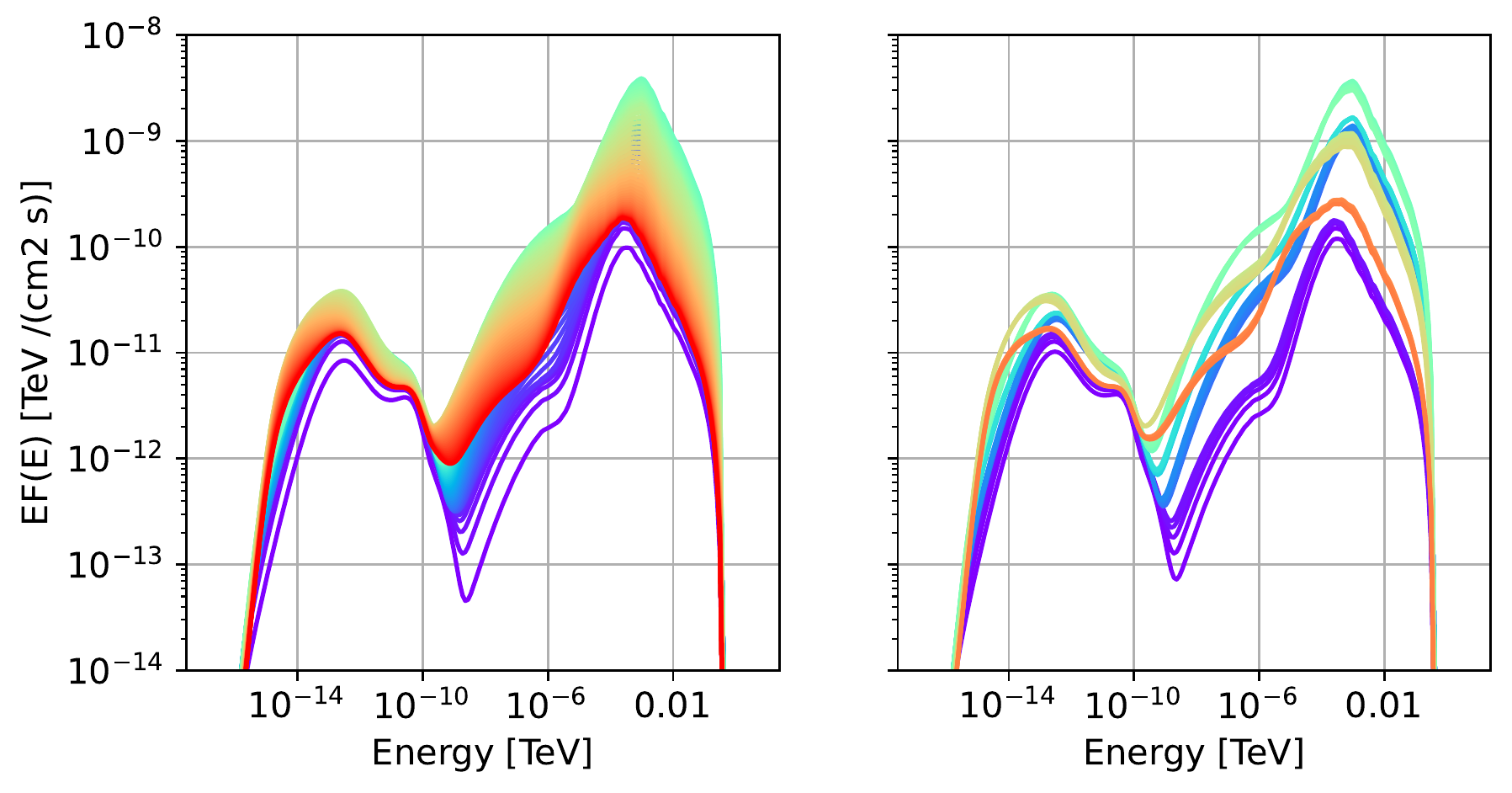}
				\caption{Leptonic model.}
				\label{fig:3C279_models_lepto}
			\end{subfigure}
			\hfill
			\begin{subfigure}[t]{0.495\textwidth}
				\includegraphics[width=\textwidth]{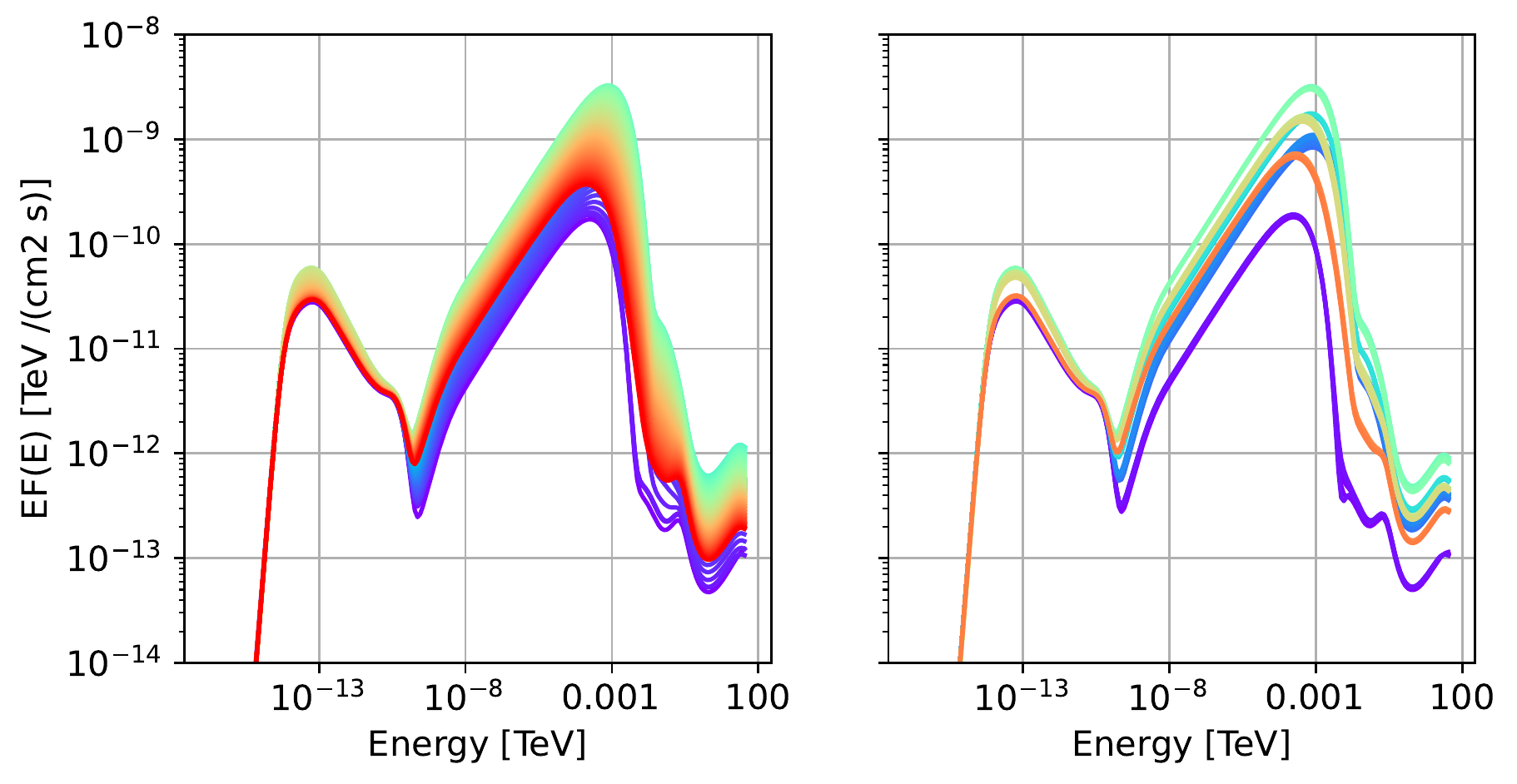}
				\caption{Hadronic model.}
				\label{fig:3C279_models_hadro}
			\end{subfigure}
			\caption{\textit{Left:} Time-dependent models used as input for the simulations of the June 2015 flare of 3C 279. Purple colors corresponding to the begining of the flare whereas red colors indicate the end of the flare. The total flare duration is $\sim$5 days. \textit{Right:} Interpolated spectra with an integration of 1 hour.}
		\end{figure}

		We use two versions of a one zone, time-dependent model as input to simulate this flare. In the leptonic scenario version, the VHE peak is due to Inverse Compton processes whereas in the hadronic scenario, proton synchrotron emission induces photon meson cascades responsible for the VHE emission. Both models are described in more details in \cite{HESS2019}. Figures \ref{fig:3C279_models_lepto} (left) and \ref{fig:3C279_models_hadro} (left) show both models at different timesteps. The different spectra are separated by 1.5 hours. From these input model snapshots, \textsc{CtaAgnVar} calculates the interpolated flux and the temporal integrated model which will be used to simulate realistic observations. The resulting interpolated spectra for an integration of 1 hour are shown in Fig. \ref{fig:3C279_models_lepto} (right) and \ref{fig:3C279_models_hadro} (right) for the letponic and the hadronic models respectively. 
		
		From these spectra, the code computes fluxes and determines the visibility and elevation angle of the source through the night. Associated with the corresponding instrumental responses, it computes simulated observations that it uses to create observed spectra as they will be seen by CTA.
These simulated spectra are then fitted with a phenomenological model chosen by the user. In the case of 3C 279, and based on the shape of the input model in the 100\,GeV--10\,TeV energy range, we use a powerlaw with an exponential cutoff for the leptonic scenario and a simple powerlaw for the hadronic scenario.
		
\begin{figure}[h]
			\centering
			\begin{subfigure}[t]{0.495\textwidth}
				\includegraphics[width=\textwidth]{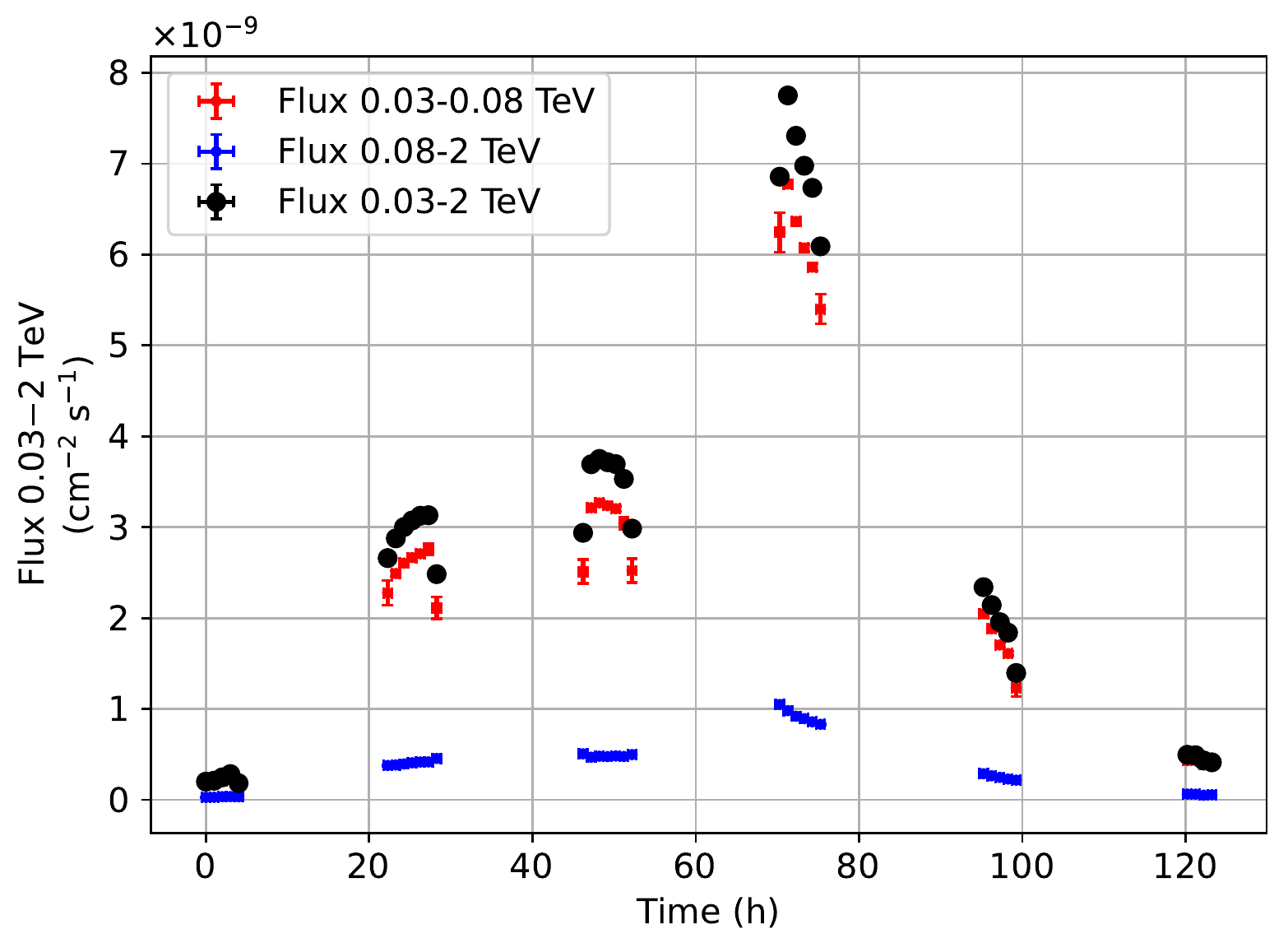}
				\caption{Leptonic model.}
			\end{subfigure}
			\hfill
			\begin{subfigure}[t]{0.495\textwidth}
				\includegraphics[width=\textwidth]{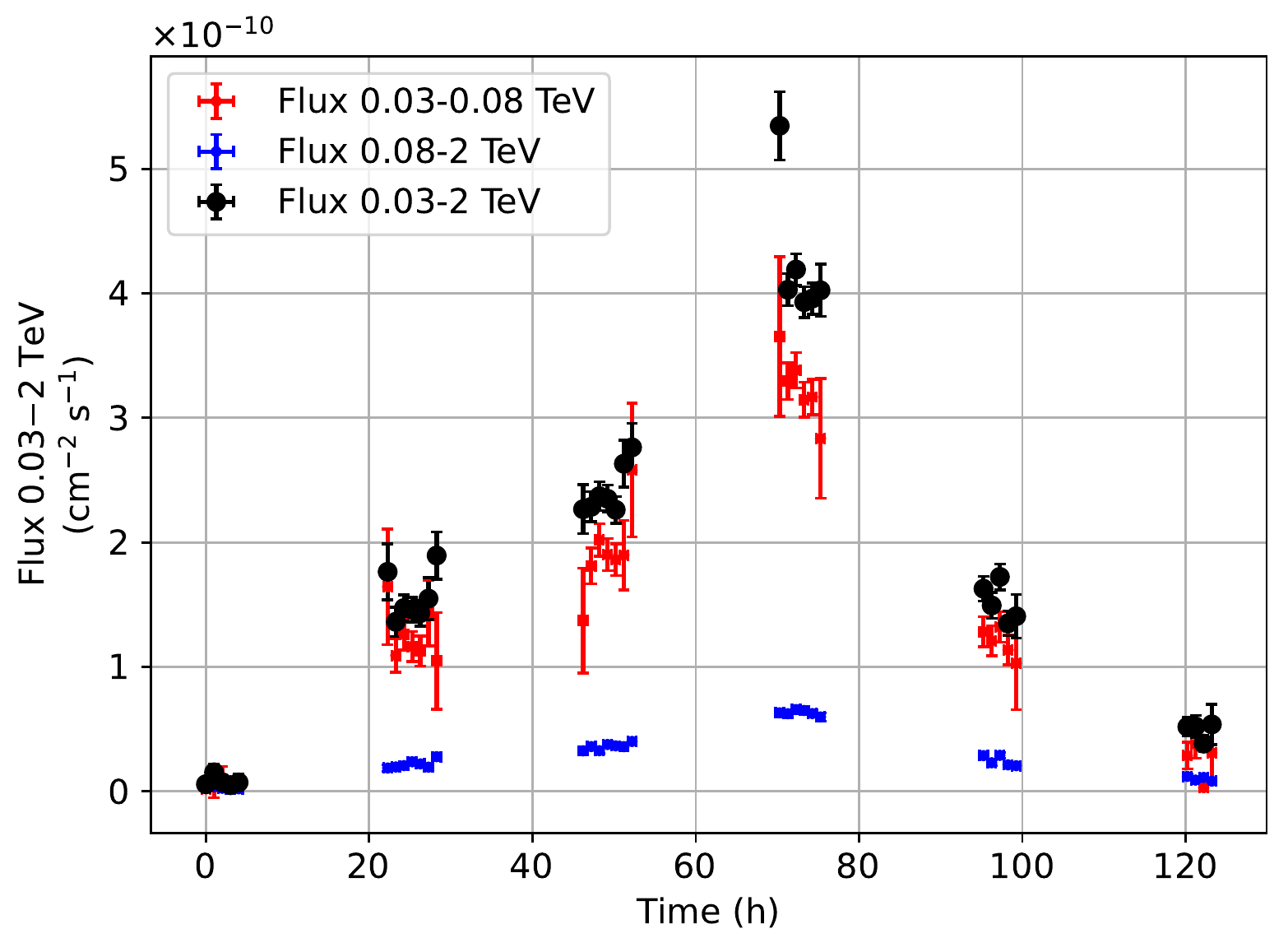}
				\caption{Hadronic model.}
			\end{subfigure}
			\caption{Simulated lightcurves of the June 2015 flare of 3C 279 with \textsc{CtaAgnVar} with a time binning of 1 hour. The different colors correspond to different energy bands; red: 0.03--0.08\,TeV, blue: 0.08--2\,TeV and black: 0.08--2\,TeV.}
			\label{fig:final_3C279}
		\end{figure}

\begin{figure}[h]
			\centering
			\begin{subfigure}[t]{0.495\textwidth}
				\includegraphics[width=\textwidth]{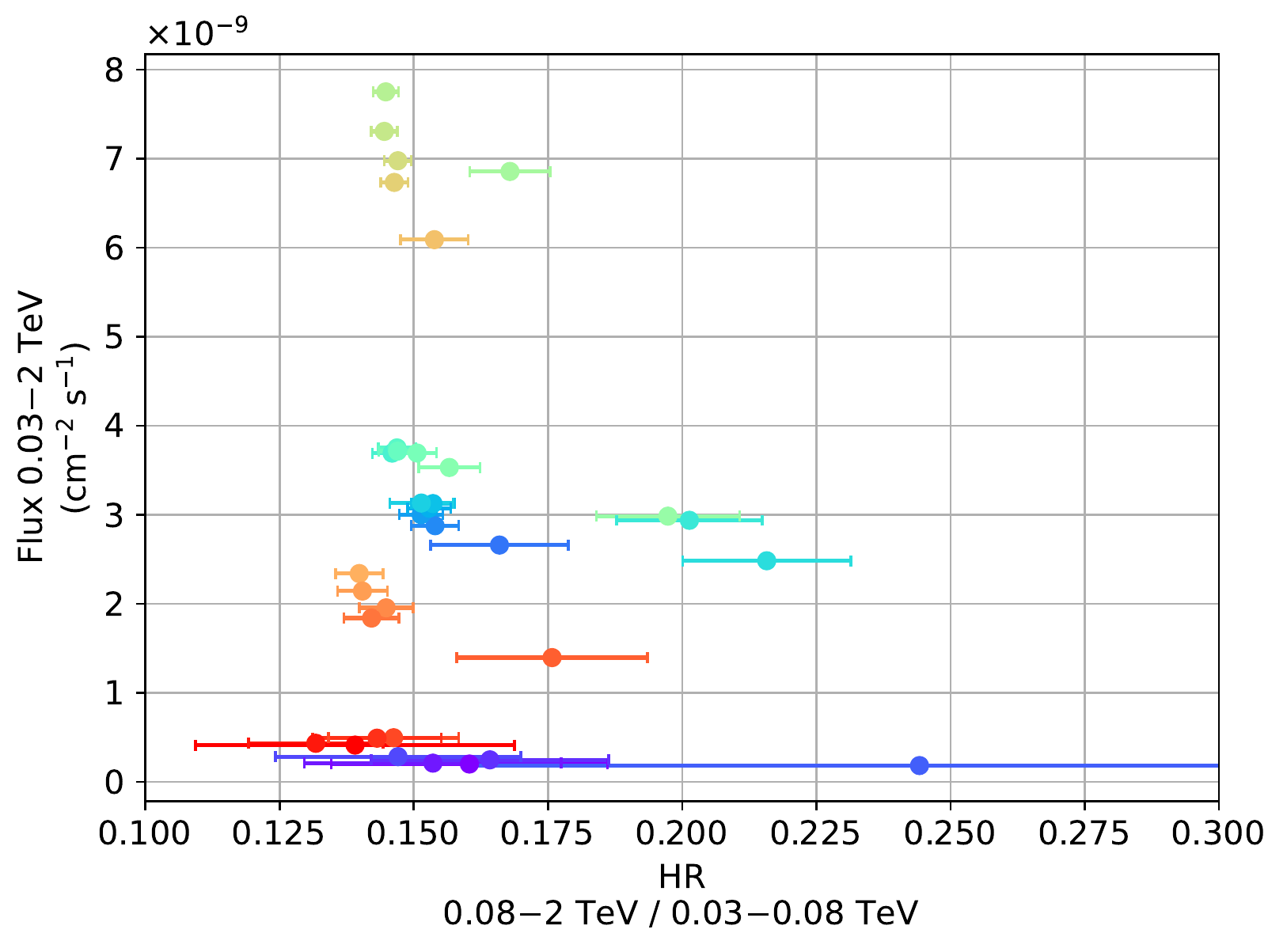}
				\caption{Leptonic model.}
			\end{subfigure}
			\hfill
			\begin{subfigure}[t]{0.495\textwidth}
				\includegraphics[width=\textwidth]{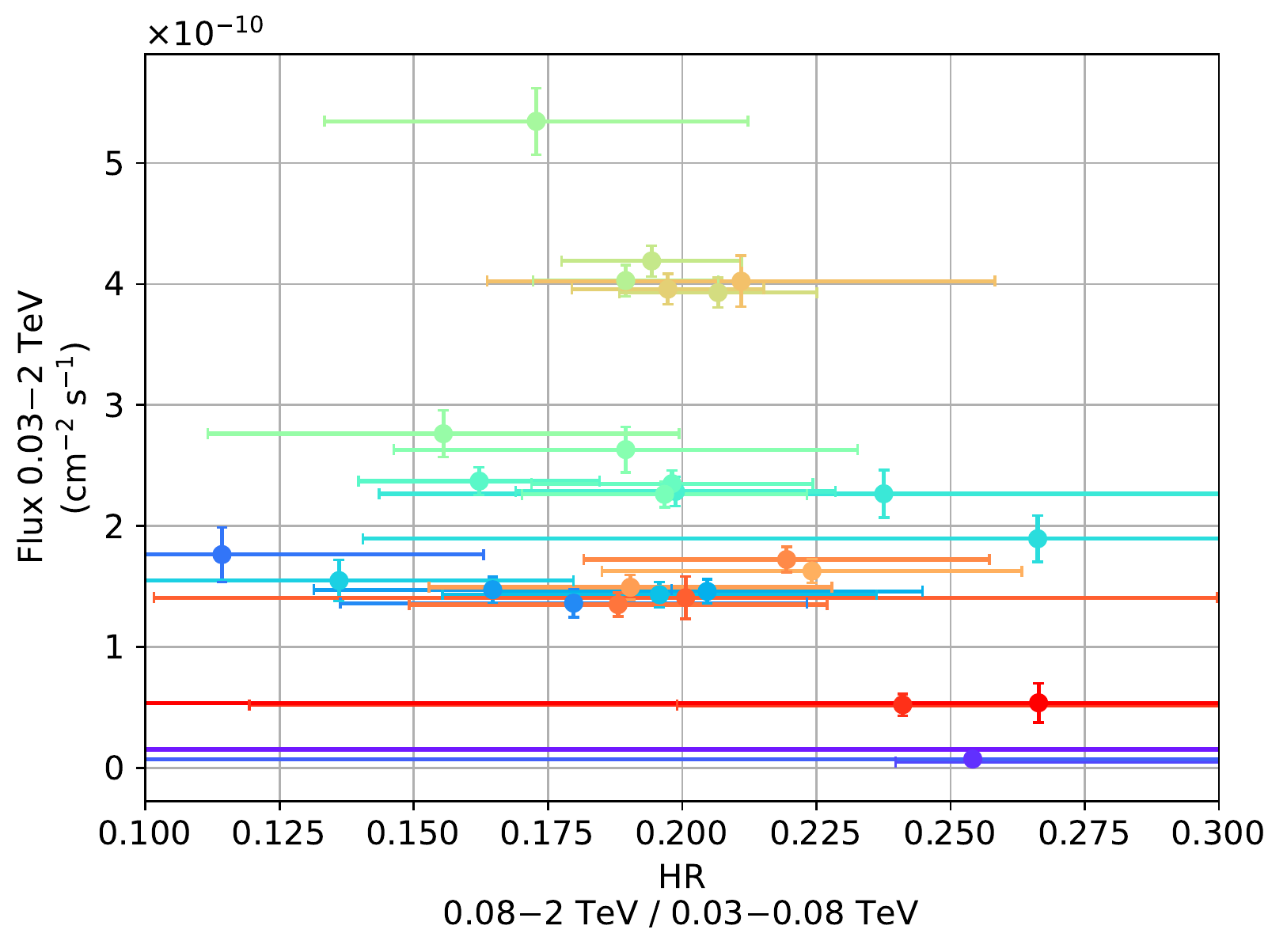}
				\caption{Hadronic model.}
			\end{subfigure}
			\caption{HIDs for the June 2015 flare of 3C 279. Different colors indicate different timestep; purple colors corresponding to the start of the flare whereas red colors corresponding to the end.}
			\label{fig:hid_3C279}
		\end{figure}

		
		Finally, the code computes fluxes for each timestep with \textit{Gammapy} methods and creates simulated observed lightcurves. Figure \ref{fig:final_3C279} shows these final lightcurves for both leptonic (left) and hadronic (right) scenarios in different energy bands and for 1 hour timestep. We also want to investigate whether CTA would be capable of giving us information on the flare evolution by observing, for instance, an hysteresis in a Hardness Intensity Diagram (HID). It is a diagnostic tool generally used in the X-ray band and which can probe the different states of heating and cooling during the flare. It represents the total flux (0.03--2\,TeV) as a function of the ratio between the high-energy band (0.08--2\,TeV) over the low energy band (0.03--0.08\,TeV). HIDs for the June 2015 flare of 3C 279 are shown on Fig. \ref{fig:hid_3C279} for both leptonic and hadronic scenarios. As expected, there is no evidence of an hysteresis for either scenario due to a constant value of the photon index and the energy of the cutoff during the flare.
		
		\subsubsection{BL Lacertae -- October 2016 flare}
\begin{figure}[h]
			\includegraphics[width=\textwidth]{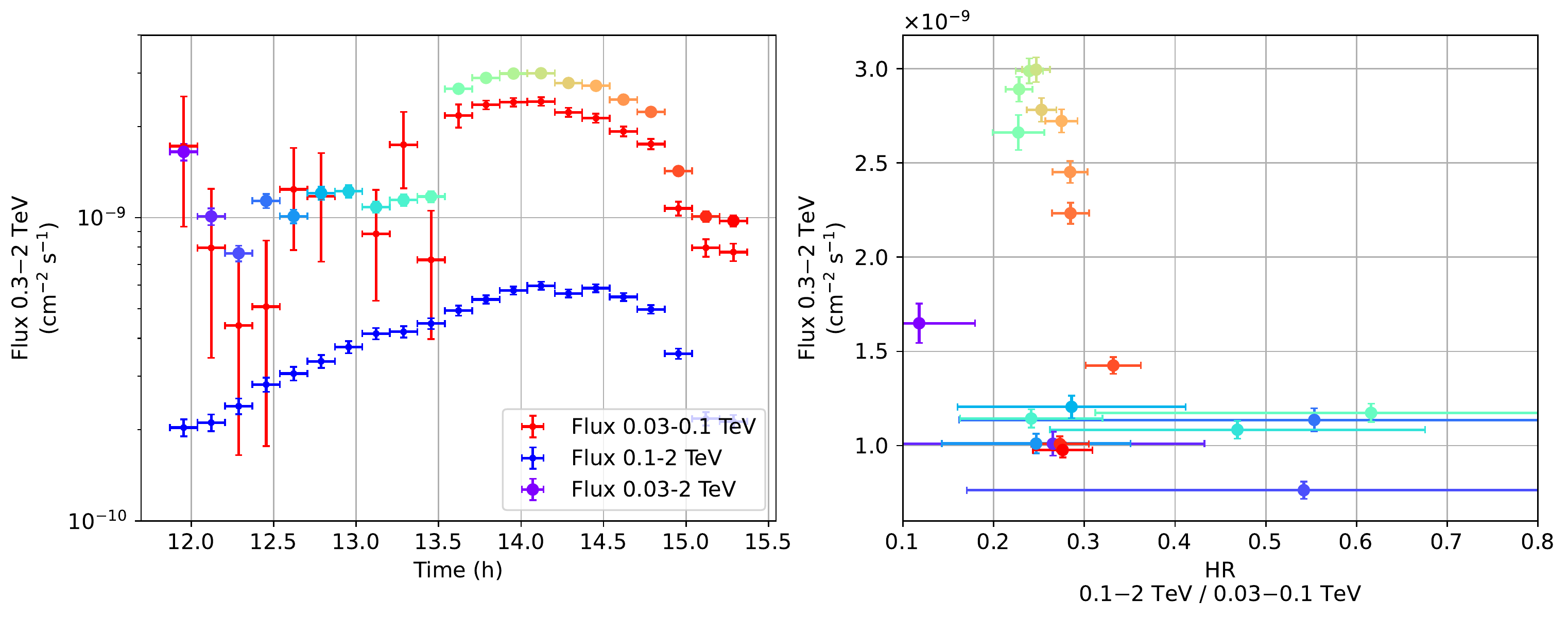}
			\caption{\textit{Left:} lightcurves of the October 2016 flare of BL Lacertae in different energy bands:  red: 0.03--0.1\,TeV, blue: 0.1--2\,TeV and rainbow: 0.1--2\,TeV. \textit{Right:} the resulting HID, the different colors indicate different times according to the lightcurve in the 0.03--2\,TeV energy bands.}
			\label{fig:BLLac_results}
		\end{figure}

		Figure \ref{fig:BLLac_results} shows the lightcurve (left) and the HID (right) for the October 2016 flare of BL Lacertae. The simulation is based on the model of \cite{Morris2019}. In this model, Synchrotron Self Compton (SSC) emission from electrons in the jets is powered by magnetic reconnection. The total duration of the flare is $\sim$10 hours, however, due to visibility and observational constrains, CTA would be able to catch only $\sim$3.5 hours of the flare. We use a timestep of 10 minutes for our simulation with \textsc{CtaAgnVar} and we fit our simulated spectra with a powerlaw with an exponential cutoff. We observe larger uncertainties during the first half of the flare; this is because the source is observed with a high zenith angle $z = 40-60$°. As a result, the first points in the HID have large uncertainties (we use the same color code in the lightcurve and the HID to show the corresponding time in both figures). We do not observe an hysteresis in the HID here because we only observe 3.5 hours of the whole flare and therefore we can only draw one part of the hysteresis. The simulation using optimal observational conditions for 10 hours shows indeed a clear hysteresis for this source (not shown in this proceedings).
		
		\subsubsection{Markarian 421 -- March 2001 flare}
\begin{figure}[h]
			\includegraphics[width=\textwidth]{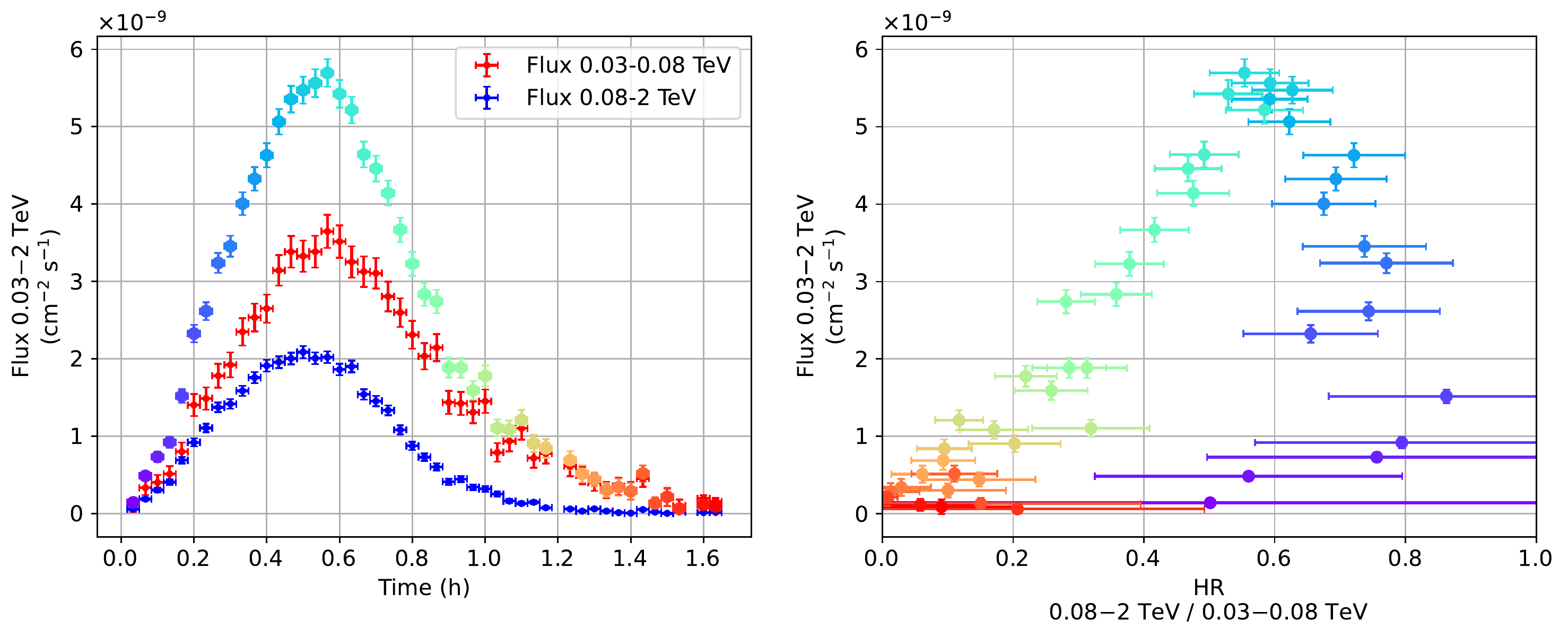}
			\caption{\textit{Left:} lightcurves of the March 2001  flare of Markarian 421 in different energy bands according to the same color code as in Fig. \ref{fig:BLLac_results}.}
			\label{fig:Mrk421_results}
		\end{figure}

		The 2001 March flare of Markarian 421 is the shortest flare simulated in this study, with a total duration of 2 hours. We use the model from \cite{Finke2008} for which the observed emission comes from SSC from electrons in the jets. Figure \ref{fig:Mrk421_results} shows the resulting simulated lightcurves (left) and HID (right) for a time binning of 2 minutes. Simulated spectra are fitted with a powerlaw with a exponential cutoff. The very short duration of the flare allows CTA to follow the whole flare without any interruptions and with zenith angles $z = 20-40$°. As a result, we observe a clear hysteresis pattern in the HID caused by the evolution of the energy cutoff during the flare.
		
	\subsection{Long-term lightcurves}
		In order to simulate long-term lightcurves, we use a different approach. As an input model, we express the flux $\phi_z$ as a function of the time $t$ and the energy $E$:
		
		\begin{equation}
		\phi_z(E,t) = e^{-\tau_{\gamma \gamma} (E, z)} \phi_0(t) \left(\frac{E}{E_0} \right)^{-\Gamma (t) - \beta ln \left(\frac{E}{E_0}\right)-\frac{E}{E_\mathrm{cut}}}
		\end{equation}
		where $E_0$ is a reference energy, $\phi_0$ is the flux of the source at $E_0$, $\Gamma$ represents the photon index of the powerlaw, $\beta$ is the curvature parameter and $E_\mathrm{cut}$ is the energy of the exponential cutoff. We also take into account the Extragalactic Background Light (EBL) absorption throught $e^{-\tau_{\gamma \gamma}}(E, z)$ which depends on the optical depth $\tau_{\gamma \gamma}(E, z)$, and the redshift of the source, $z$.
		
\begin{figure}[h]
			\includegraphics[width=\textwidth]{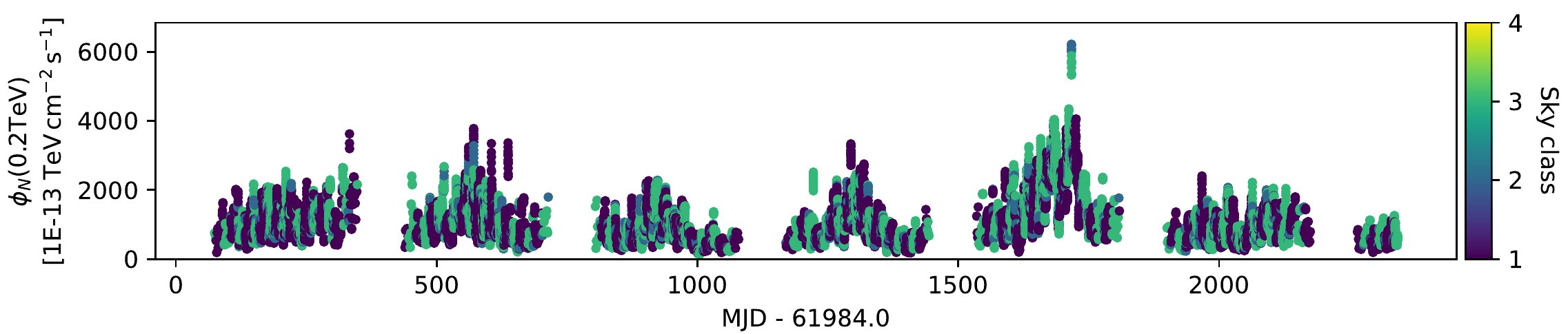}
			\caption{Long-term lightcurve simulated for Markarian 421. Different colors indicate different "Sky class": dark time with optimal observational conditions (dark blue), grey time with higher Night Sky Background  (blue), bright time which corresponds to day light (green). Yellow colors are not shown on this plot and correspond to the source being not visible by CTA.}
			\label{fig:longterm_lc}
		\end{figure}

		In this model, parameters $E_\mathrm{cut}$, $\beta$ and $z$ are time-independent and their value for 14 AGN of dedicated CTA Key Science Project (KSP) are obtained from \cite{ctaconsortium2020}. Temporal variability is injected through $\phi_0(t)$ and $\Gamma (t)$; $\phi_0(t)$ follows a log-normal process with a colored-noise time dependence as described in \cite{Emmanoulopoulos2013} and $\Gamma (t)$ is correlated with $\phi_0(t)$ and follows the "harder when brighter" relation.
		As a result, we can compute simulated time-dependent spectra that we use to create long-term lightcurves as shown on Fig. \ref{fig:longterm_lc} for Markarian 421. The color code corresponds to different "Sky class" (described in the caption). Simulations have been done for the 14 KSP sources. In the future, these will also serve as an input for \textsc{CtaAgnVar} in order to define the best observational strategy for CTA long-term observations.
		
\section{Summary and perspectives}

In this work, we have developed new tools to simulate AGN flares and long-term lightcurves. For very short, bright flares (< 3.5 hours), CTA will be able to follow the whole flare without any interruption and with a very fine time-binning.  For longer flares, it will be able to catch part of the flare and eventually reconstruct an hysteresis in a HID. This preliminary study shows that CTA will offer new possibilities to exploit time-resolved analysis and to probe AGN short and long-term variability. This work is in progess. Up to now we have analysed 3 typical flares with 4 theoretical models but more theoretical models will be investigated and compared. We also plan to reconstruct power spectra and duty cycles of simulated long-term lightcurves with \textsc{CtaAgnVar}. All these results will be reported in detail in a future CTA consortium paper.

\end{document}